\documentclass[5p]{elsarticle}

\usepackage{xcolor}
\usepackage{hyperref}
\usepackage[T1]{fontenc}
\usepackage[utf8]{inputenc}
\usepackage{textcomp} 
\usepackage{lmodern} 
\usepackage{multirow}
\usepackage{amsmath}
\usepackage{amsfonts}
\usepackage[switch]{lineno} 
	
\graphicspath{{./figures/}}
\biboptions{sort&compress}

\hypersetup{
    breaklinks = true,
    colorlinks = true,
    pdftitle = {}, 
    pdfauthor = {},
    pdfkeywords = {},
    linkcolor = blue,
    citecolor = blue,
    filecolor = black,
    urlcolor = magenta
}

\makeatletter 
\def\@author#1{\g@addto@macro\elsauthors{\normalsize%
    \def\baselinestretch{1}%
    \upshape\authorsep#1\unskip\textsuperscript{%
      \ifx\@fnmark\@empty\else\unskip\sep\@fnmark\let\sep=,\fi
      \ifx\@corref\@empty\else\unskip\sep\@corref\let\sep=,\fi
      }%
    \def\authorsep{\unskip,\space}%
    \global\let\@fnmark\@empty
    \global\let\@corref\@empty  
    \global\let\sep\@empty}%
    \@eadauthor={#1}
}
\makeatother

\definecolor{notecolor}{RGB}{0,0,191}
\definecolor{warningcolor}{RGB}{191,0,0}
\definecolor{donecolor}{RGB}{0,127,0}
\definecolor{todocolor}{RGB}{191,100,0}
\newcommand{\etal}{\textit{et al.}} 
\newcommand{\g}{$\mathrm{\Gamma}$} 
\newcommand{\mub}{$\mu_\mathrm{B}$}

\begin{document}
\begin{sloppypar} 

\title{Exchange Interactions and Magnon Spectra in bcc Fe(001) Thin Films:\\ A Computational Study
}

\author{Joanna Marciniak\corref{cor1}$^{1,2}$}
\ead{joanna.marciniak@ifmpan.poznan.pl}
\author{Wojciech Marciniak$^{1,3}$}
\author{Jorge Luis Briseño-Gómez$^{4}$}
\author{Jan Rusz$^{1}$}

\cortext[cor1]{Corresponding author}
\address{$^{1}$Department of Physics and Astronomy, Uppsala University, P.O. Box 516, 75120, Uppsala, Sweden}
\address{$^{2}$Institute of Molecular Physics, Polish Academy of Sciences, M. Smoluchowskiego 17, 60-179 Poznań, Poland}
\address{$^{3}$Institute of Physics, Poznań University of Technology, Piotrowo 3, 60-965, Poznań, Poland}
\address{$^{4}$Departamento de Física, Facultad de Ciencias, Universidad Nacional Autónoma de México, Ciudad Universitaria, Av. Universidad 3000, Mexico City, 04510, Mexico}

\begin{abstract}

We combined density functional theory with atomistic spin dynamics to understand how surfaces influence magnon dispersion. 
For this purpose, we investigate thin layers of body-centered cubic iron and compare them with the bulk material.
We present how the broken symmetry affects exchange interactions and magnon behavior.

Our results reveal enhanced magnetic moments and exchange interactions for surface atoms compared to bulk Fe. 
The increase in exchange interaction integrals is mostly visible for first- and second-nearest neighbors.
We also find a significant spread in exchange values within each coordination shell. 
Magnon dispersion analysis shows, for the layered system, pronounced band blurring, a significant reduction of maximum frequencies relative to the previously reported bulk values, and the emergence of a low-frequency band outside the \g{}-point. 
Our study shows that relying on bulk exchange parameters introduces errors in the magnon spectra of ultrathin systems, underscoring the need for a detailed, system-specific approach.

\end{abstract}


\maketitle

\section{Introduction}

Magnonics\,\textemdash{}\,processing information via magnetic moments instead of electric current\,\textemdash{}\,promises lower costs and higher operating speeds than conventional electronics~\cite{barman_2021_2021,flebus_2024_2024}. 
Understanding the phenomena that govern magnon behavior is crucial for achieving these advantages.
One topic of interest in device engineering is understanding how defects influence magnon behavior.
Defects appear broadly in devices.
We focus on a specific example\,\textemdash{}\,interfaces and surfaces (abrupt crystal discontinuities).

We follow a good practice in choosing a starting point for this type of research, and use a simple system based on a well-known material.
In magnetism, one of the best-known systems are bcc Fe, hcp Co and fcc Ni.
Thin films of these materials are widely studied, both experimentally (e.g.~\cite{scheurer_magnetic_1993,kipferl_stabilization_2005,chuang_magnetic_2014}) and computationally (e.g.~\cite{szunyogh_initio_1999,mankovsky_electricfield_2021,donascimento_confined_2024}).
Out of these systems, we choose the former\,\textemdash{}\,bcc Fe.

One key parameter for computational investigations of spin dynamics is the nearest-neighbor exchange integral, which defines spin interactions. 
From a computational point of view, exchange interactions in layered bcc Fe are investigated from the perspective of ultrathin films deposited on experimentally accessible substrates, e.g., W~\cite{bergqvist_atomistic_2013a,zakeri_probing_2016}, Ir~\cite{chuang_magnetic_2014}, Rh~\cite{meng_direct_2014}, Cu~\cite{bergqvist_atomistic_2013a}, or Pt~\cite{mankovsky_electricfield_2021}.
Free-standing or cross-cut bcc Fe films have been investigated less commonly.
Wang~\etal{}\ looked at nearest-neighbor exchange integrals in a 3-monolayer-thick slab of bcc Fe(001) in vacuum~\cite{wang_spinwave_2016}.
Mankovsky~\etal {} studied exchange-integral values for more coordination shells using the spin-polarized relativistic Korringa-Kohn-Rostoker Green function calculation method, but did so for a free-standing Fe monolayer only~\cite{mankovsky_electricfield_2021}.
In contrast, do Nascimento~\etal{} simulated magnon dispersions for free-standing Fe films with nm-scale thicknesses using bulk-like exchange integrals~\cite{donascimento_confined_2024}.
To the best of our knowledge, the literature does not yet provide a full multi-shell exchange tensor for a free-standing bcc Fe(001) slab across several thicknesses.


In this work, we examined how the material-vacuum surface influences exchange-interaction integrals between atoms near the surface and within ultrathin films, and how the picture changes with increasing thickness.
For this purpose, we used density functional theory (DFT) calculations and the maximally localized Wannier functions approach on the test system, bcc Fe, in ultrathin films with thicknesses of 10, 14, and 20 monolayers.
Using the full set of exchange-interaction integrals within the chosen radius, we then simulated the magnon dispersion in the selected films.
The magnon dispersions were simulated using atomistic spin dynamics (ASD) based on the Landau-Lifshitz-Gilbert equation~\cite{eriksson_atomistic_2017}.
This allowed us to investigate how film thickness influences magnon dispersion.

\section{Computational details}

Etz~\etal{}~\cite{etz_atomistic_2015} concluded that atomistic spin dynamics is a powerful tool for investigating surface-magnon problems.
Especially when connected with \textit{ab initio} calculations of ground-state parameters, it allows prediction of material-specific behaviors.
We applied such a DFT-ASD approach in our studies.

\begin{figure}[t]
\includegraphics[width=0.75\columnwidth]{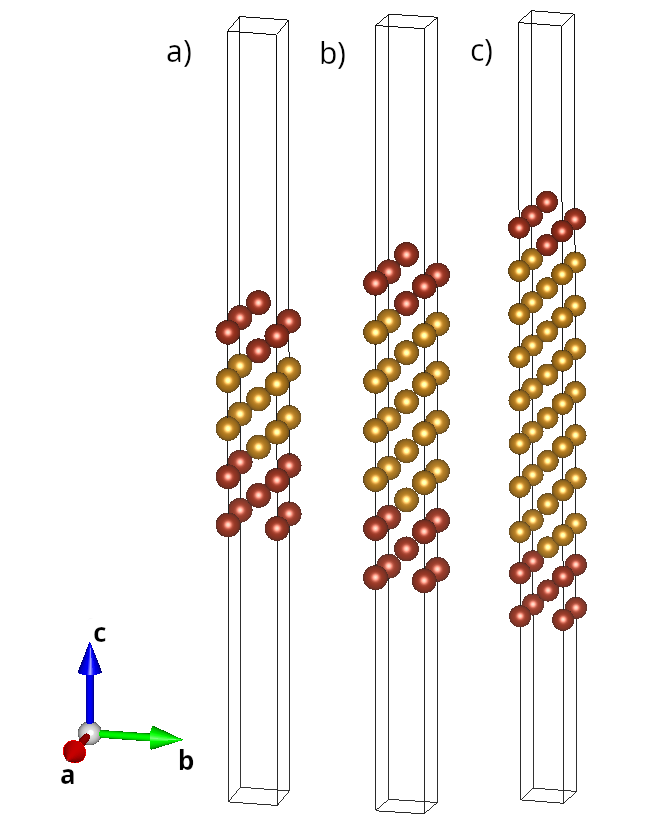}
\caption{Computational unit cells for (a) 10-, (b) 14-, and (c) 20-monolayer-thick films as used for DFT calculations with the SIESTA code. Three Fe atoms nearest to the surface with prolonged $4s$ orbitals are red. We prepared the structure visualization using VESTA~\cite {momma_VESTA_2008}.}
\label{fig-structures}
\end{figure}

As a model system for our investigations, we used free-standing ultrathin films of bcc Fe.
This allowed us to examine the influence of broken crystal symmetry without additional interactions with another material.
We examined three thicknesses: 10 (1.43~nm), 14 (2.002~nm), and 20 (2.86~nm) atomic monolayers.
We used a tetragonal unit cell in the \textit{P\,4} space group.
The geometry of the films was not optimized.
We wanted to separate the effects of the reduced coordination number of atoms at the surface, without introducing effects stemming from the change of interatomic distances.
To set the computational cell size, we used the experimental lattice constant $a = b = 2.86$~\AA{} as the base size and $c = 45.76$~\AA{} for 10- and 14-monolayer-thick films, and $c = 51.48$~\AA{} for a 20-monolayer-thick film as the computational cell height.
We then filled this computational cell with an appropriate number of cubic cells of bcc Fe with a commensurate lattice parameter (2.86~\AA{}).
This procedure provides more than 20~\AA{} of vacuum between periodic images of the slab under investigation, preventing interactions between them.
Figure~\ref{fig-structures} shows the computational unit cells used in DFT calculations.
Red-colored atoms indicate atoms with expanded $4s$ orbital basis functions (see below).

For DFT calculations, we used SIESTA 5.0.2~\cite{soler_siesta_2002,garcia_siesta_2020} with a fully relativistic pseudopotential provided in the general gradient approximation Perdew-Burke-Ernzerhof (GGA-PBE) exchange-correlation functional from DojoProject~\cite{hamann_optimized_2013, vansetten_pseudodojo_2018, garcia_psml_2018}.
Calculations were provided for 32x32x1 (10- and 14-monolayer thick films) and 36x36x1 (20-monolayer thick film) \textbf{k}-point meshes with spin-orbit coupling included and tolerances set at 10$^{-7}$ and 10$^{-5}$~eV for density matrix and Hamiltonian accuracies, respectively.
The lack of neighbors near the surface reduces the number of functions describing electron behavior there, which can lead to an inappropriate description of the electronic structure. 
We used the simple method described by Garcia-Gil \etal{}~\cite{garcia-gil_optimal_2009} to avoid introducing artificial atoms in subsequent exchange-interaction calculations with the TB2J package.
Based on the optimized basis set, we optimize the cut-off radius for the $4s$ orbital for the three atoms closest to the surface.
The optimized cut-off radius for the $4s$ orbital is 9.4486~Bohr radii.
Next, we calculate maximally localized Wannier functions from SIESTA results using built-in functions.
We used the resulting Wannier functions to compute the full set of exchange-interaction integrals for the 7x7x1 supercell with the TB2J Python package~\cite{he_tb2j_2021}.


Finally, we performed an ASD simulation using the UppASD code~\cite{antropov_abinitio_1995,skubic_method_2008,etz_atomistic_2015}.
We reduced the computed set of exchange interaction integrals to interactions within a radius of around 18\,\AA{}.
Simulations were performed for 40x40x1 supercells composed of the DFT computational cells, with periodic boundary conditions in the plane of the films.
We performed the simulations at 10\,K.
We used 4000 snapshots collected every 50 timesteps, with a timestep of 10$^{-16}$\,s.
There was assumed 0.5 T magnetic field in direction perpendicular to the surface and no anisotropy.
We set the Gilbert damping parameter to 0.0165, following~\cite {barati_gilbert_2014a}, assuming a scattering rate \g{} = 0.01\,eV.

\section{Results and Discussion}

\begin{table}[t]
\caption{Spin ($m_s$) and orbital ($m_l$) magnetic moments calculated for 10-, 14-, and 20-monolayer-thick films of bcc Fe (001) using SIESTA 5.0.2~\cite{soler_siesta_2002,garcia_siesta_2020}. Magnetic moments from DFT calculations are oriented along $c$ direction.}
\label{tab-mag-moms}
\center
\begin{tabular}{c | c c | c c | c c}
        \hline
        \hline
        &\multicolumn{2}{|c}{10 atom slab} & \multicolumn{2}{|c}{14 atom slab}& \multicolumn{2}{|c}{20 atom slab}\\
        No. & $m_s$ & $m_l$ & $m_s$ & $m_l$ & $m_s$ & $m_l$ \\
        \hline
        1  & 3.072 & 0.182 & 3.076 & 0.185 & 3.078 & 0.187 \\
        2  & 2.222 & 0.114 & 2.223 & 0.116 & 2.233 & 0.117 \\
        3  & 2.289 & 0.098 & 2.303 & 0.097 & 2.302 & 0.098 \\
        4  & 2.117 & 0.092 & 2.143 & 0.094 & 2.135 & 0.094 \\
        5  & 2.144 & 0.088 & 2.184 & 0.092 & 2.173 & 0.091 \\
        6  & 2.144 & 0.088 & 2.159 & 0.092 & 2.153 & 0.091 \\
        7  & 2.117 & 0.092 & 2.143 & 0.092 & 2.157 & 0.092 \\
        8  & 2.289 & 0.098 & 2.143 & 0.092 & 2.150 & 0.092 \\
        9  & 2.222 & 0.114 & 2.159 & 0.092 & 2.162 & 0.092 \\
        10 & 3.072 & 0.182 & 2.184 & 0.092 & 2.176 & 0.093 \\
        11 & -- & -- & 2.143 & 0.094 & 2.176 & 0.093 \\
        12 & -- & -- & 2.303 & 0.097 & 2.162 & 0.092 \\
        13 & -- & -- & 2.223 & 0.116 & 2.150 & 0.092 \\
        14 & -- & -- & 3.076 & 0.185 & 2.157 & 0.092 \\
        15 & -- & -- & -- & -- & 2.153 & 0.091 \\
        16 & -- & -- & -- & -- & 2.173 & 0.091 \\
        17 & -- & -- & -- & -- & 2.135 & 0.094 \\
        18 & -- & -- & -- & -- & 2.302 & 0.098 \\
        19 & -- & -- & -- & -- & 2.233 & 0.117 \\
        20 & -- & -- & -- & -- & 3.078 & 0.187 \\
        \hline
        \hline
    \end{tabular}
\end{table}

In the first step of the implemented procedure, we obtained layer-resolved magnetic moment values, separated into spin and orbital components.
Table~\ref{tab-mag-moms} presents the exact values. 
The total magnetic moments for the center atomic layers of the films in the range 2.232\textendash{}2.269\,\mub{} are close to the reported bulk values (2.19\textendash{}2.25\,\mub{})~\cite{chapman_machinelearned_2022}.
Also, the magnetic moment enhancement obtained in the surface and close-to-surface layers agrees well with reported values by Ohnishi \etal{}~\cite{ohnishi_surface_1983} (2.98\,\mub{}), Autes \etal{}~\cite{autes_magnetism_2006} (spin magnetic moment: $\sim 3.14$\,\mub{}, orbital magnetic moment: $\sim 0.155$\,\mub{}) and monolayer calculations of Mankovsky \etal{}~\cite{mankovsky_electricfield_2021} (3.013\,\mub{}).
This magnetic moment enhancement is more pronounced for orbital magnetic moments, which are around twice as large at the surface as at the center of the films.
For spin magnetic moments, we observe above a 40\% enhancement at the surface compared to the central layers.

We also observe oscillations in the spin magnetic moment across the film thickness in all structures.
Similar behavior has been reported not only in bcc Fe films~\cite{ohnishi_surface_1983,autes_magnetism_2006,meixner_magnetic_2024}, but also in other magnetic thin films~\cite{marciniak_l10_2023,marciniak_magnetic_2024,marciniak_first-principles_2024}. 
In the thickest, 20-monolayer film, orbital magnetic moments also show slight oscillations in the middle of the film, starting after the fourth layer from the surface.
This agrees with results reported by Autes \etal{}~\cite{autes_magnetism_2006}.

\begin{figure*}[t]
\includegraphics[width=0.67\columnwidth]{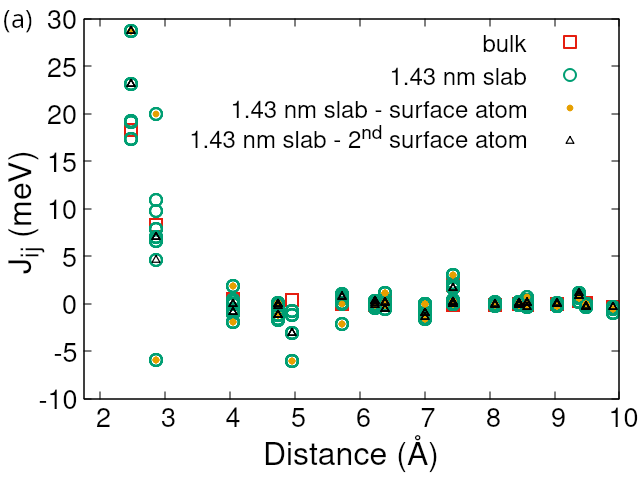} \hspace{0.2cm}
\includegraphics[width=0.67\columnwidth]{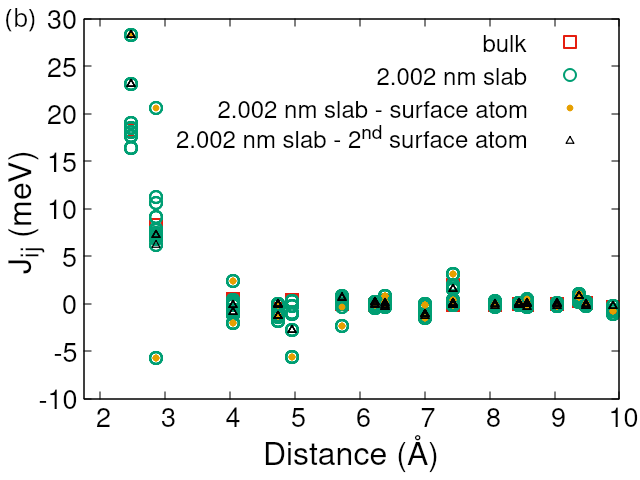} \hspace{0.2cm}
\includegraphics[width=0.67\columnwidth]{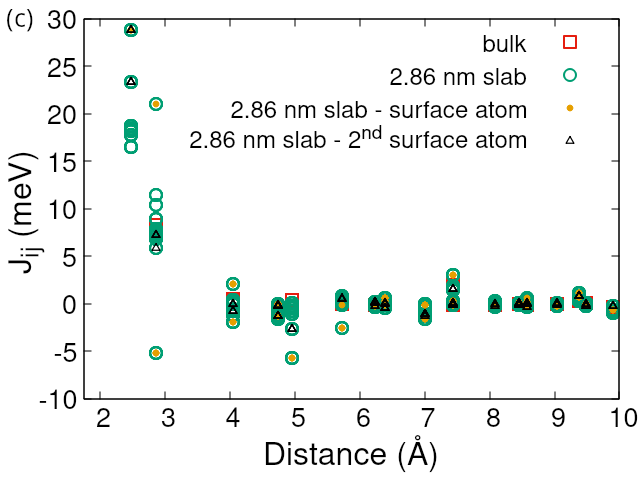}
\caption{Comparison of obtained exchange interaction integrals for bcc Fe(001) thin films of (a) 10, (b) 14, and (c) 20 monolayers thickness with bulk bcc Fe values. Bulk values are marked by red squares and slab by green empty circles. Surface and second-surface atom exchange interaction integrals are shown as yellow full circles and black triangles, respectively.}
\label{fig-exchange_integrals}
\end{figure*}

Next, we calculated exchange interaction integrals for the bulk and for all film thicknesses.
Figure~\ref{fig-exchange_integrals} shows the results.
A few visible deviations from the bulk exchange interaction integrals appear.
First, the first- and second-nearest-neighbor interaction integrals show a significant enhancement.
Similar enhancements have been reported in previous works for different thicknesses and for systems either on the substrate~\cite{bergqvist_atomistic_2013a,meng_direct_2014,chuang_magnetic_2014,zakeri_probing_2016,mankovsky_electricfield_2021} or free-standing~\cite{mankovsky_electricfield_2021}.

At the same time, this enhancement is accompanied by a spread of values, especially for first- and second-nearest-neighbor interactions.
Most of the deviation occurs in the exchange-interaction integrals with the surface atom, followed by interactions with the second layer closest to the surface.
This spread in second-nearest-neighbor interactions yields antiferromagnetic exchange-interaction integrals, consistent with previous observations~\cite{bergqvist_atomistic_2013a,meng_direct_2014,zakeri_probing_2016,mankovsky_electricfield_2021}.
Interestingly, the differences between the examined thicknesses are small.

Finally, we performed ASD simulations using the calculated exchange interaction integrals and obtained magnon dispersions for all thicknesses.
We observe magnon frequency maxima in 21\textendash{}23 THz range around the M high-symmetry point, depending on film thickness.
The maximum magnon frequency increases with film thickness, consistent with Kipferl \etal{}~\cite{kipferl_stabilization_2005} observations that the frequency increases with Fe film thickness on Au.
The maximum frequency agrees well with Chuang \etal{}~\cite{chuang_magnetic_2014} observations for Fe films on Ir(001).
In contrast, Buczek \etal{}~\cite{buczek_different_2011} and do Nascimento \etal{}~\cite{donascimento_confined_2024} reported significantly higher magnon frequencies, much closer to bulk bcc Fe magnon frequencies.
The differences seem to arise mostly from the method used to obtain the dispersions.
In \cite{buczek_different_2011}, magnon dispersion was obtained from time-dependent DFT.
Although the method is more accurate, they studied only the Fe(100) monolayer both \textit{in vacuo} and on Cu(100), which may significantly alter the result.
In contrast, do Nascimento \etal{}~\cite{donascimento_confined_2024} used a Hamiltonian diagonalization approach with a linear spin-wave approximation to obtain magnon dispersions, but their result is based on bulk bcc Fe exchange-interaction integrals.
As we show in Figure~\ref{fig-exchange_integrals}, the exchange interaction integrals differ significantly between thin-film systems and the bulk structure.
Kipferl \etal{}~\cite{kipferl_stabilization_2005} experimentally showed that in bcc Fe(100) thin films on Au, a decrease in exchange interaction integrals increases magnon frequencies.

The magnon band widens near the M high-symmetry point due to reduced dimensionality and may arise from multiple bands with very small spacing.
The higher magnon band has a shape similar to that presented by Buczek \etal{}~\cite{buczek_different_2011}, but different from the dispersion presented by do Nascimento \etal{}~\cite{donascimento_confined_2024}.
Additionally, we observe a low-frequency band that goes to 0 frequency on the X\textendash{}M and M\textendash{}\g{} paths.
This band decreases in intensity as slab thickness increases.

\begin{figure*}[t]
\center
\includegraphics[width=0.79\textwidth]{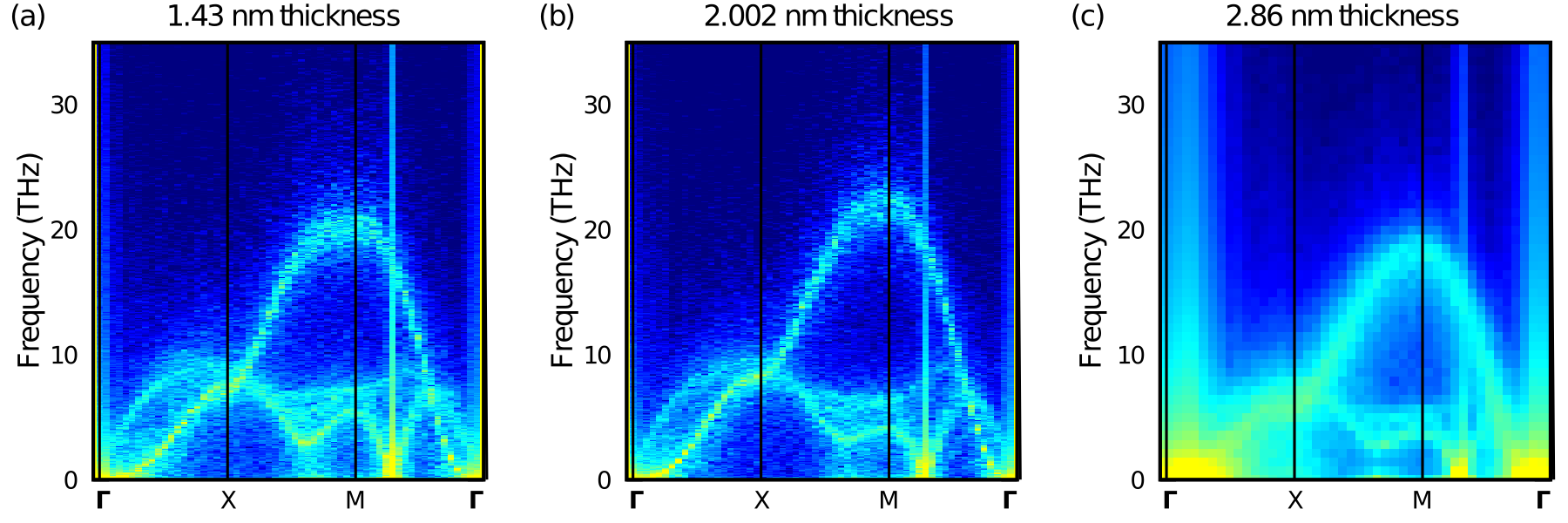}
\caption{Magnon dispersions for bcc Fe(001) thin films with thicknesses of (a) 10, (b) 14, and (c) 20 monolayers, corresponding to 1.43, 2.002, and 2.86 nm, respectively.}
\label{fig-magnon_dispertions}
\end{figure*}

\section{Summary and Conclusions}

In summary, we performed a multi-code workflow from DFT computations to ASD simulations for bcc Fe(001) thin films with 10-, 14-, and 20-monolayer thicknesses.
We obtained a full set of exchange interaction integrals for the studied systems.
The exchange integrals show an increase in the slab compared with bcc Fe bulk calculations.
This increase in the first- and second-neighbor interactions for the surface atom is above 50\% and 30\%, respectively.
We showed that exchange interaction integrals for each neighborhood shell exhibit a spread of values, up to 45\% of the maximum value for nearest-neighbor interactions.
In magnon dispersions, we mostly observe band blurring and maximum frequencies reduced to about 5 times lower than bulk values.
Additionally, a lower-frequency band appears for thin slabs, with frequency approaching 0 outside \g{}-points.
This band decreases in intensity for the thickest investigated slab.
Overall, we show that using a complete set of exchange interactions for an ultrathin system calculated for the low-dimensional system can improve results over using bulk exchange-interaction values.

\section*{Acknowledgements}
We acknowledge support from the Swedish Research Council (grant no.~2025-0451), the Olle Engkvist Foundation (grant no.~214-0331), and the Knut and Alice Wallenberg Foundation (grant no.~2022.0079). The simulations were enabled by resources provided by the National Academic Infrastructure for Supercomputing in Sweden (NAISS) at NSC Centre, partially funded by the Swedish Research Council through grant agreement no.~2022-06725. J.~L.~B.-G. acknowledges financial support from the \textit{Secretaría de Ciencia, Humanidades, Tecnología e Innovación} (SECIHTI), Mexico, in the form of a PhD scholarship. Additional support was provided by the \textit{Apoyo a Estancias de Larga Duración} grant from the \textit{Programa de Apoyo a Estudiantes de Posgrado} of the \textit{Posgrado en Ciencias Físicas}, \textit{Universidad Nacional Autónoma de México}.

\bibliography{references}

\end{sloppypar}
\end{document}